\documentclass[12pt]{article}

\def\d{\displaystyle}
\def\be{\begin{equation}}
\def\ee{\end{equation}}

\title{\bf Primordial matter density contrast and the size of the very early
  universe in the
  Quantum Big Bang theory of the cosmological constant}
\author{Budh Ram \\
Physics Department \\
New Mexico State University\\ 
Las Cruces, New Mexico 88003, USA\\ 
 \\
 and \\
 \\
Prabhu-Umrao Institute of Fundamental Research \\
A2/214 Janak Puri, New Delhi 110058, India \\
}
\date{}

\begin{document}

\maketitle

\begin{abstract} 
We calculate the amount of primordial matter
density contrast and the size of the very early universe in the recent
Quantum Big Bang theory [Arxiv:0705.4549[gr-qc](2007)] of the 
cosmological constant.  We obtain
$(\delta\rho/\rho)_M = 1.75 \times 10^{-5}$, {\it without} the
introduction of an adjustable free parameter.  Harrison-Zel'dovich
$k$-dependence with $A = 64/9\pi^2 = 0.72$ and $n = 1$ in
$|\delta_k|^2 = Ak^n$ arises inherently.  The size of the universe
with which it enters the classical Friedmann-Robertson-Walker (FRW)
phase comes out to be 0.2 cm.  We conclude that the hypothesis of
classical inflation at an early stage of cosmic evolution is {\bf not}
needed. 
\end{abstract}

\newpage

In a recent paper [1] we constructed the Quantum Big Bang
theory of the cosmological constant in which its presently observed
very small value $\sqrt{\Lambda/3} = 1.00 \times 10^{-61}$ --
appears naturally.  (Throughout this paper we use units in which $G =
c = \hbar = 1$).  The Quantum Big Bang theory [1] is constructed by
applying the method [2] earlier given by Ram, Ram and Ram to the classical
de Sitter line element.  In the resulting quantum theory there appear
discrete de Sitter zero-point quantum states with zero-point energies
$\d{3\over2} \omega_0, \d{3\over2} \omega_1, \d{3\over2} \omega_2,
\cdots; \omega_0, \omega_1, \omega_2, \cdots$ being the frequencies of
{\it independent} oscillators.
\bigskip

As a result of the first quantum bang (QB) the zero-point energy
$\d{3\over2} \omega_0$ is reduced to $\d{3\over2} \omega_1$ and the
difference $\d{3\over2} (\omega_0-\omega_1) = 2(1)\omega_1$ is
transformed into one pair of mass quanta, each quantum of frequency
$\omega_1$.  As a result of the 2nd QB, the zero-point energy
$\d{3\over2} \omega_1$ is further reduced to $\d{3\over2}\omega_2$,
the difference $\d{3\over2} (\omega_1 - \omega_2) = 2(2)\omega_2$
being converted to 2 pairs of mass quanta, each mass quantum of
frequency $\omega_2$.  And so on, till as a result of the last 43-rd
QB, 43 pairs of mass quanta, each of frequency $\omega_{43}$ are created.
\bigskip

Everyone of these mass quanta contributes to the matter density
$\rho_M$, the contribution of each quantum being proportional to the
{\it square} of its frequency $\omega$ (see [1]).
\bigskip

First we wish to calculate the matter density contrast
$(\delta\rho/\rho)_M$ given in terms of the frequency contrasts
$\d{\Delta \omega_n \over \omega_n} \ (n = 1,2,\cdots,43)$ by the
relation 
\be
\left({\delta \rho \over \rho}\right)_M = \sum^{43}_{n=1}
2\left({\Delta \omega_n \over \omega_n}\right)
\label{one}
\ee
since the frequencies $\omega_1,\omega_2,\cdots,\omega_{43}$ contribute
{\it independently} to the matter density contrast.
\bigskip

In order to do that, let us look closely at the conceptually
mathematical nature of a QB.
\bigskip

After the application in [1] of the method [2] to the de Sitter line
element the resulting quantum equation $(\omega = \sqrt{\Lambda/3})$
\be
\left(-{1\over2} {d^2 \over dr^2} + {1\over2} \omega^2 r^2\right)U =
{1 \over 8\pi} U
\label{two}
\ee
has only one eigenvalue $\epsilon = \d{1 \over 8\pi}$ in
\be
\epsilon = \left(2n + {3\over2}\right) \omega, \ n = 0,1,2,\cdots .
\label{three}
\ee
Assigning $\epsilon = \d{1 \over 8\pi}$ to the $n = 0$ ground state
results in an oscillator of frequency $\omega_0 = \d{1 \over 12\pi}$.
On the other hand, assigning this eigenvalue $\d{1 \over 8\pi}$ to the
$n = 1$ state results in an oscillator of a different frequency
$\omega_1 = \d{1 \over 28\pi}$, the relation between $\omega_0$ and
$\omega_1$ being
\be
{3\over2} \omega_0 = \left(2(1) + {3\over2}\right) \omega_1.
\label{four}
\ee
Eq.(\ref{four}) is interpreted as: the first QB transforms the energy
difference $\d{3\over2} (\omega_0 - \omega_1) = 2(1) \omega_1$ to one
pair of mass quanta, each mass quantum of frequency $\omega_1$.  In
quantum language, the first QB is transition of the isotropic
oscillator of frequency $\omega_1$, from its $|n=1\rangle$ state to
its $|n=0\rangle$ ground state and spontaneous emission (a la
Einstein) of one $p$-{\it quantum} (paired-quantum) of frequency
$\omega_1$.  This is equivalent to a {\it Planck} oscillator
spontaneously emitting one quantum of frequency $\omega_1$, and is
thus analogous to an atomic transition from one quantum state to
another with spontaneous emission of a light quantum of frequency
$\omega_1$ (see [3] and references cited therein).
\bigskip

Hence one can calculate the spread $\Delta \omega$ in the frequency
$\omega$ of the 
spontaneously emitted mass-quantum in the same manner one calculates
the spread in the frequency of the light-quantum in the atomic case,
namely, by using the famous {\it Golden Rule} of time-dependent
perturbation theory [4].  Thus  
\be
\Delta \omega = \omega_{AB} = \int 2\pi \rho_\omega|\langle
B|H_I|A\rangle|^2 d\Omega.
\label{five}
\ee
In Eq.(\ref{five}), $\omega_{AB}$ is the total transition probability
per unit time for spontaneous emission of a mass quantum of frequency
$\omega$ for the isotropic oscillator transition $|A\rangle
\rightarrow |B\rangle$, 
\be
\rho_\omega = {\omega^2 \over (2\pi)^3}
\label{six}
\ee
is the density of states in the $\omega$ continuum with the emitted
quantum of momentum $\vec k$ ($k = \omega$) directed within the solid
angle $d\Omega$, and $H_I$ is the self-interaction of the oscillator
which we take to be
\be
H_I = p_r = {dr \over d\tau} \ (\tau \ {\rm is \ the \
proper \ time, \ see [1]})
\label{seven}
\ee
analogous to the atomic case.
\bigskip

For the interaction (\ref{seven}), the squared matrix element
\be
|\langle B|H_I|A\rangle|^2 = \omega^2 |\langle B|r|A\rangle|^2.
\label{eight}
\ee
Then the contribution $\left(\d{\delta\rho \over \rho}\right)_n$ from
the $n$-th QB to the total matter density contrast is simply given by 
\be
\left({\delta\rho \over \rho}\right)_n = \left({2\Delta \omega_n \over
\omega_n}\right) = {2\omega^3_n \over \pi} |\langle 0|r|n\rangle|^2.
\label{nine}
\ee
For a given $n$ the matrix element $|\langle 0|r|n\rangle|^2$ can be
easily calculated using the isotropic oscillator wave functions given
in [5] (we explicitly reproduce the first three wave functions in the
Appendix).  For the first QB
\be
|\langle 0|r|1\rangle|^2 = {2 \over 3\pi \omega_1}
\label{ten}
\ee
giving
\be
\left({\delta\rho \over \rho}\right)_1 = {4\omega^2_1 \over 3\pi^2} =
1.75 \times 10^{-5}
\label{eleven}
\ee
with $\omega_1 = \d{1 \over 28\pi}$ (see [1]).
\bigskip

For the second QB
\be
|\langle 0|r|2\rangle|^2 = {1 \over 30\pi\omega_2}
\label{twelve}
\ee
giving
\be
\left({\delta\rho \over \rho}\right)_2 = {\omega^2_2 \over 15\pi^2} =
7 \times 10^{-8}
\label{thirteen}
\ee
with $\omega_2 = 3.1 \times 10^{-3}$ (see [1]).
\bigskip

Contribution from the 3rd QB is of order $10^{-11}$, i.e., negligible,
and it is expected to be negligible for the remainder of the QB's in
view of how the $\omega_n$'s decrease with increasing $n$.  Thus the
major contribution to the density contrast comes from the first QB and
$(\delta\rho/\rho)_M$ remains almost constant at the value $\simeq
1.75 \times 10^{-5}$ during the successive 41 QB's.  And it enters the
classical FRW phase with this value.
\bigskip

With $\omega_1 = k_1 \rightarrow k$, Eq.(\ref{eleven}) is rewritten as
\be
\left({\delta\rho \over \rho}\right)^2 = {16 \over 9\pi^4} k^3 k.
\label{fourteen}
\ee
Note the Harrison-Zel'dovich $k$-dependence in (\ref{fourteen}).
Equating (\ref{fourteen}) with Eq. (7.80) of [6] (see also [7]), namely 
\be
\left\langle \left({\delta M \over M}\right)^2_R\right\rangle = {A \over
2} \left({k^3 k \over 2\pi^2}\right)_{k=R^{-1}},
\label{fifteen}
\ee
one gets for the Zel'dovich amplitude $A$ in $|\delta_k|^2 = Ak^n \ (n=1)$
the value [8]
\be
A = {64 \over 9\pi^2} = 0.72.
\label{sixteen}
\ee

In essence the matter density contrast $\left(\d{\delta\rho \over
\rho}\right)_M \simeq 1.75 \times 10^{-5}$, with its
Harrison-Zel'dovich $k$-dependence [9] and $A = 0.72$, results from
the Quantum Big Bang.  And then the classical FRW description takes
over.

Next we answer the question: what is the size of the universe when it
enters the classical FRW phase?

The universe enters the classical FRW description after the {\bf last}
43-rd QB with zero-point energy $\d{3\over2} \omega_{43} = 1.5 \times
10^{-61}$ (see [1]).  The state of the universe with this zero-point
energy is described by the radial quantum equation [1]
\be
\left[-{1 \over 2r^2} {\partial \over \partial r} \left(r^2 {\partial
\over \partial r}\right) + {1\over2} \omega^2_{43} r^2\right]\psi_{43}
= \left(2(43) + {3\over2}\right)\omega_{43} \psi_{43}
\label{seventeen}
\ee
with a completely spherically symmetric state function $\psi_{43}$.
Thus the size of the universe with which it enters the classical FRW
phase, $a_{ECP}$, is simply $\sqrt{4\pi}$ times [1,10] the square
root of the expectation value of $r^2$ for the state $\psi_{43}$,
viz., 
\be
a_{ECP} = \sqrt{4\pi \langle r^2 \rangle_{43}}.
\label{eighteen}
\ee
The expectation value of $r^2$ for the state $\psi_{43}$ is most
easily calculable by using the well-known result [11] that for a given
quantum state of the isotropic oscillator the expectation value of the
potential energy equals half the total energy:
\be
\left\langle {1\over2} \omega^2_{43} r^2 \right\rangle_{43} =
{1\over2} \left(2(43) + {3\over2}\right)\omega_{43}
\label{nineteen}
\ee
which, with $\omega_{43} = 10^{-61}$ (see [1]), gives
\be
\sqrt{\langle r^2\rangle_{43}} = 3 \times 10^{31} (= 4.8 \times 10^{-2}
\ {\rm cm}).
\label{twenty} 
\ee
Or
\be
a_{ECP} = 0.2 \ {\rm cm}.
\label{twentyone}
\ee
Thus the universe enters the classical FRW phase with a size equal to
0.2 cm.

With the present size of the universe $\simeq 10^{28}$ cm and the
present cosmic microwave background (CMB) radiation temperature
$\simeq 2.4 \times 10^{-4}$ eV (2.73 
K), the temperature and entropy of the very early universe come out to
be $10^{16}$ GeV and $10^{87}$, respectively.

It is clear from the results obtained in this paper that the
hypothesis of classical inflation [12] at an early stage of cosmic
evolution is {\bf not} needed. 

\bigskip\bigskip
\begin{center}
{\bf Acknowledgments}
\end{center}
\bigskip

The author thanks Rajeev Bhalerao, Arun Ram, Nilam Ram, Narayan
Banerjee, Subha Majumdar, and Anatoly Klypin for useful conversations,
the Tata Institute of Fundamental Research, Mumbai for a pleasant
stay, and the Mathematics Department, Melbourne University, Australia
for a delightful visit.  He is grateful to Rajeev Bhalerao for his
continued help and support.

\newpage
\begin{center}
{\bf Appendix}
\end{center}
\bigskip

Here we reproduce the first three radial wave functions $R_{n,\ell=0}
(r), \ n = 0,1,2$, for the isotropic oscillator from [5].
\begin{eqnarray}
R_0 (r) &=& \pi^{-1/4} \alpha^{3/2} 2 e^{-{1\over2}(\alpha r)^2}, \ \alpha =
\omega^{1/2}, \nonumber \\[2mm]
R_1 (r) &=& \pi^{-1/4} \alpha^{3/2} {2^{3/2} \over 3^{1/2}}
\left({3\over2} - (\alpha r)^2\right) e^{-{1\over2}(\alpha r)^2}, \nonumber
\\[2mm] R_2(r) &=& \pi^{-1/4} \alpha^{3/2} {2^{3/2} \over (15)^{1/2}}
\left[{15 \over 4} - 5(\alpha r)^2 + (\alpha r)^4\right]
e^{-{1\over2}(\alpha r)^2}. \nonumber
\end{eqnarray}
\bigskip\bigskip

\begin{center}
{\bf References and Notes}
\end{center}
\bigskip

\begin{enumerate}
\item[{[1]}] Budh Ram, ArXiv:0705.4549 [gr-qc] (2007).  
\item[{[2]}] B. Ram, A. Ram, and N. Ram, ArXiv:gr-qc/0504030 (2005);
B. Ram, Phys. Lett. \underbar{A265}, 1 (2000).  See also, B. Ram and
J. Shirley, ArXiv:gr-qc/0604074 (2006).
\item[{[3]}] B. Ram, ArXiv:quant-ph/0503109 (2005).
\item[{[4]}] W. Heitler, {\it The Quantum Theory of Radiation},
third edition (Dover Publications, New York 1984), \S17 and 18.
\item[{[5]}] M.A. Preston and R.K. Bhaduri, {\it Structure of the
Nucleus} (Addison-Wesley Publishing Co. 1975), p.120.
\item[{[6]}] J.V. Narlikar, {\it An Introduction to Cosmology},
3rd edition (Cambridge Univ. Press 2002), p.258.
\item[{[7]}] T. Padmanabhan, {\it Structure Formation in the
Universe} (Cambridge Univ. Press 1993), Chap.5.
\item[{[8]}] The best fit for the power-law LCDM model for
WMAP + 2d FGRS + Ly$\alpha$ data gives $A \simeq 0.75$; see D.N. Spergel
{\it et al.}, Ap J. Supplement Series \underbar{148}, 175 (2003).
\item[{[9]}] The Harrison-Zel'dovich model is consistent with the WMAP
three-year data; see W.H. Kinney, E.W. Kolb, A. Melchiorri, and
A. Riotto, ArXiv:astro-ph/0605338 (2006).
\item[{[10]}] H.P. Robertson, Phil. Mag. {\bf 5}, 835 (1928).
\item[{[11]}] L.I. Schiff, {\it Quantum Mechanics}, 3rd edition
(McGraw-Hill, New York, 1955), p.180.
\item[{[12]}] A.H. Guth, Phys. Rev. {\bf D23}, 347 (1981).
\end{enumerate}     

\end{document}